\documentclass[journal]{IEEEtran}
\usepackage{color,array,amsthm}
\usepackage{cite}
\usepackage{amsmath,amssymb,amsfonts,graphicx}
\usepackage{algorithmic}
\usepackage{svg}
\usepackage{graphicx}
\usepackage{mathtools}
\usepackage{textcomp}
\usepackage{xcolor}
\usepackage{textcomp}
\graphicspath{ {./} }

\usepackage[caption=false,font=footnotesize]{subfig}
\def\BibTeX{{\rm B\kern-.05em{\sc i\kern-.025em b}\kern-.08em
    T\kern-.1667em\lower.7ex\hbox{E}\kern-.125emX}}
\begin{document}
\title{Dilated POCS: Minimax Convex Optimization}
\author{Albert R. Yu, \IEEEmembership{Member, IEEE}, Robert J. Marks II, \IEEEmembership{Fellow, IEEE}, Keith E. Schubert, \IEEEmembership{Senior Member, IEEE}, Charles Baylis, Austin Egbert, \IEEEmembership{Member, IEEE}, \IEEEmembership{Senior Member, IEEE}, Adam Goad, \IEEEmembership{Student Member, IEEE}, Sam Haug
}

\maketitle

\begin{abstract}
\textit{Alternating projection onto convex sets} (POCS) provides an iterative procedure to find a signal that satisfies two or more convex constraints when the sets intersect. For nonintersecting constraints, the method of simultaneous projections produces a minimum mean square error  (MMSE) solution. In certain cases, a minimax solution is more desirable. Generating a minimax solution is possible using {\em dilated POCS}.   The minimax solution uses morphological dilation of nonintersecting signal convex constraints.  The sets are progressively dilated to the point where there is intersection at a minimax solution. Examples are given contrasting the MMSE and minimax solutions in problems of tomographic reconstruction of images. Dilated POCS adds a new imaging modality for image synthesis.
		Lastly, morphological erosion of signal sets is suggested as a method to shrink the overlap when sets intersect at more than one point.

\end{abstract}


\section{Introduction}
\label{sec:introduction}
\IEEEPARstart{A}{lternating} \textit{projection onto convex sets} (POCS)
\cite{Jiang,stark1998,marks2009,youla1982} is a popular method for convex minimum mean square error (MMSE) minimization. We propose POCS optimization based on morphological dilation that provides a minimax solution as an alternative to  MMSE optimization. 
 
 POCS requires a nonempty intersection of convex constraints in order to arrive at a fixed point solution; otherwise a greedy limit cycle is reached \cite{breg1965, gubin1967}. A solution to the non-intersecting sets can use the method of simultaneous weighted projections to arrive at a compromise solution. The result is a (possibly weighted) MMSE solution \cite{Combettes1997,marks2009,stark1998}.
	
	An alternative to the minimum mean squares error solution is the minimax solution. Here a solution is found such that the maximum Euclidean distance to each convex set constraint is minimized. The approach uses morphological dilation of convex sets of signals. Convex sets of signals are dilated until there is overlap. Like the MMSE approach, each constraint can be weighted. The MMSE and minimax solutions are contrasted with examples from linear inversion and tomography.
	

\section{Background}

 Based on the pioneering work of Youla, Webb \cite{youla1982}, Sezan and Stark \cite{ sezan1982, Sezan1984,stark1998}, 
 POCS applications to medical imaging has become of singular importance. Applications include 
     limited angle tomographic reconstruction \cite{Jaffe}, time delay tomography \cite{Haik,Trumbo}, X-ray CT image construction \cite{Feichtinger,Huang,Kim, Liu},
    low dosage cardiac prescreening \cite{Mandal}, 
    biomagnetic imaging \cite{Ramon},
    and
        fan-beam \cite{Peng} \& cone-beam \cite{Yu} tomography.
    Applications in MRI \cite{Samsonov} include cardiovascular imaging 
    \cite {Hsu} spin-echo MRI \cite{Riek},
    2-D multislice MRI \cite{Shilling}  and
    MRI artifact suppression \cite{Weerasinghe}. 
 
    POCS has also been applied to numerous other fields including signal recovery \cite{peng1989}, artificial neural networks \cite{sezan1990, yeh1991, marks1987}, sensor network source location \cite{Hero},  image processing \cite{Park2004} (e.g. image ghost \cite{Lee} \&  blur \cite{Ozkan} correction),  graph matching \cite{vanWyk}, geophysics \cite{Abma,Gan,Menke}, holography \cite{Jennison},
    and time-frequency analysis \cite{oh1990, oh1994}.

	Conventional POCS used in these papers uses an MMSE solution or variations thereof. The introduction of dilated POCS offers the alternative of a minimax error solution. The method does not claim to always offer superior performance, but is offered as an alternative modality for image synthesis.

	\section{POCS}
	
	Before presenting the minimax dilated POCS solution, a brief overview of POCS is needed for purposes of completeness, continuity and establishment of notation.  In-depth tutorials are available from Stark \& Yang \cite{stark1998} and Marks \cite{marks2009} (Chapter 11).

	A set $C$ is convex if for all $\vec{x}\in C$ and all $\vec{y}\in C$ then $\alpha \vec{x} + (1-\alpha) \vec{y}\in C$ for all $0\leq \alpha \leq 1$. The projection onto the convex set $C$ of $\vec{z}$ is the unique point in $C$ closest to $\vec{z}$ in the mean-square (Euclidian) sense. The unique orthogonal projection onto set $C$ is denoted by the equation
	$$ \vec{x} = P _{C} ( \vec{z}) $$ 
	and means for a given $\vec{z}$,
	$$ \vec{x} = \arg \min_{\vec{y} \in C} \|  \vec{z} -\vec{y}  \| .$$
	
\begin{figure}[!t]
		\centering
		\includegraphics[width=3in]{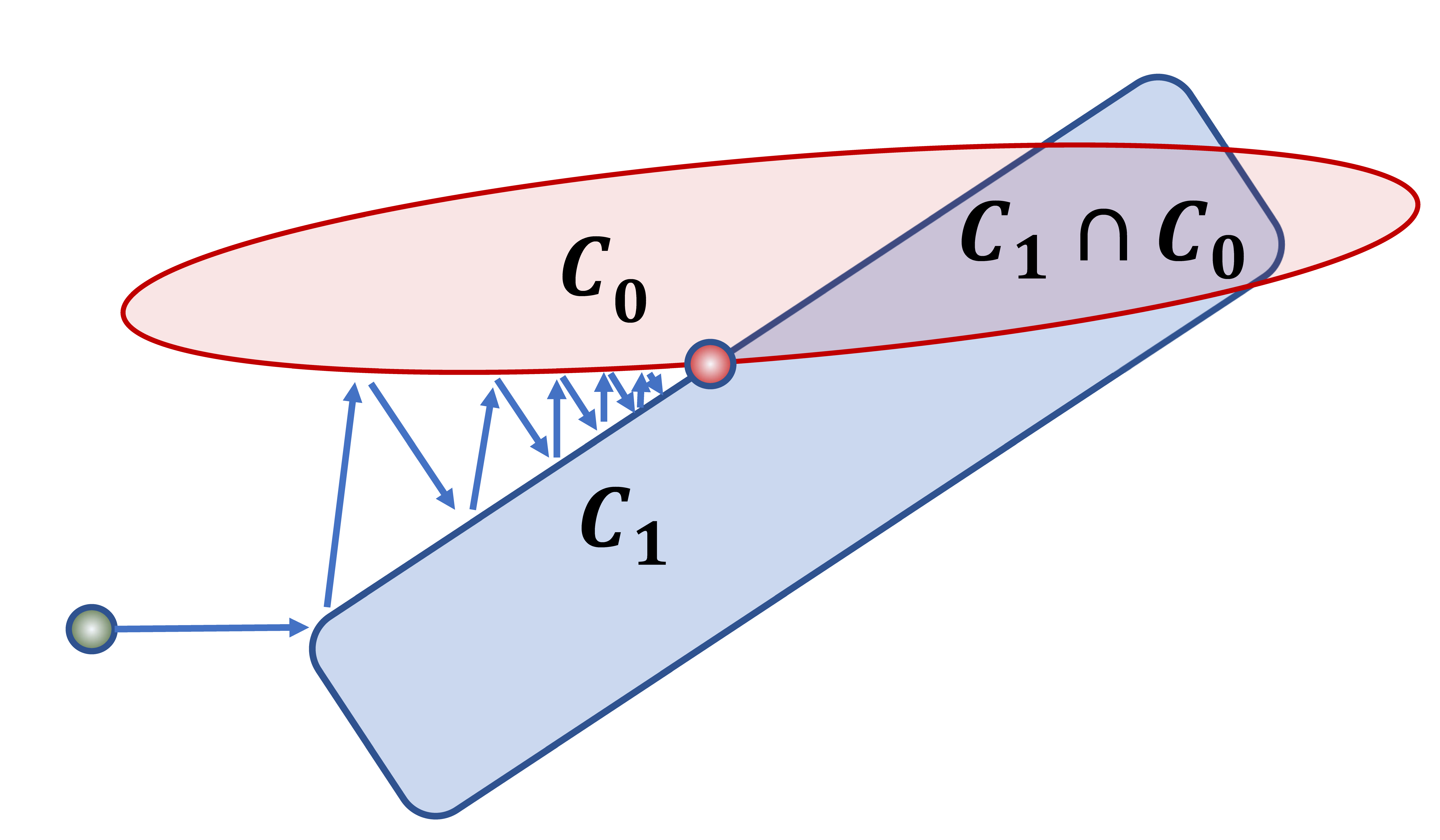}
		\caption{An illustration of the convergence of alternating projections. POCS will converge onto a point on the intersection of convex constraint sets from any random initialization.}
		\label{fig_MAP}
		\vspace{5mm}
		\includegraphics[width=2.5in]{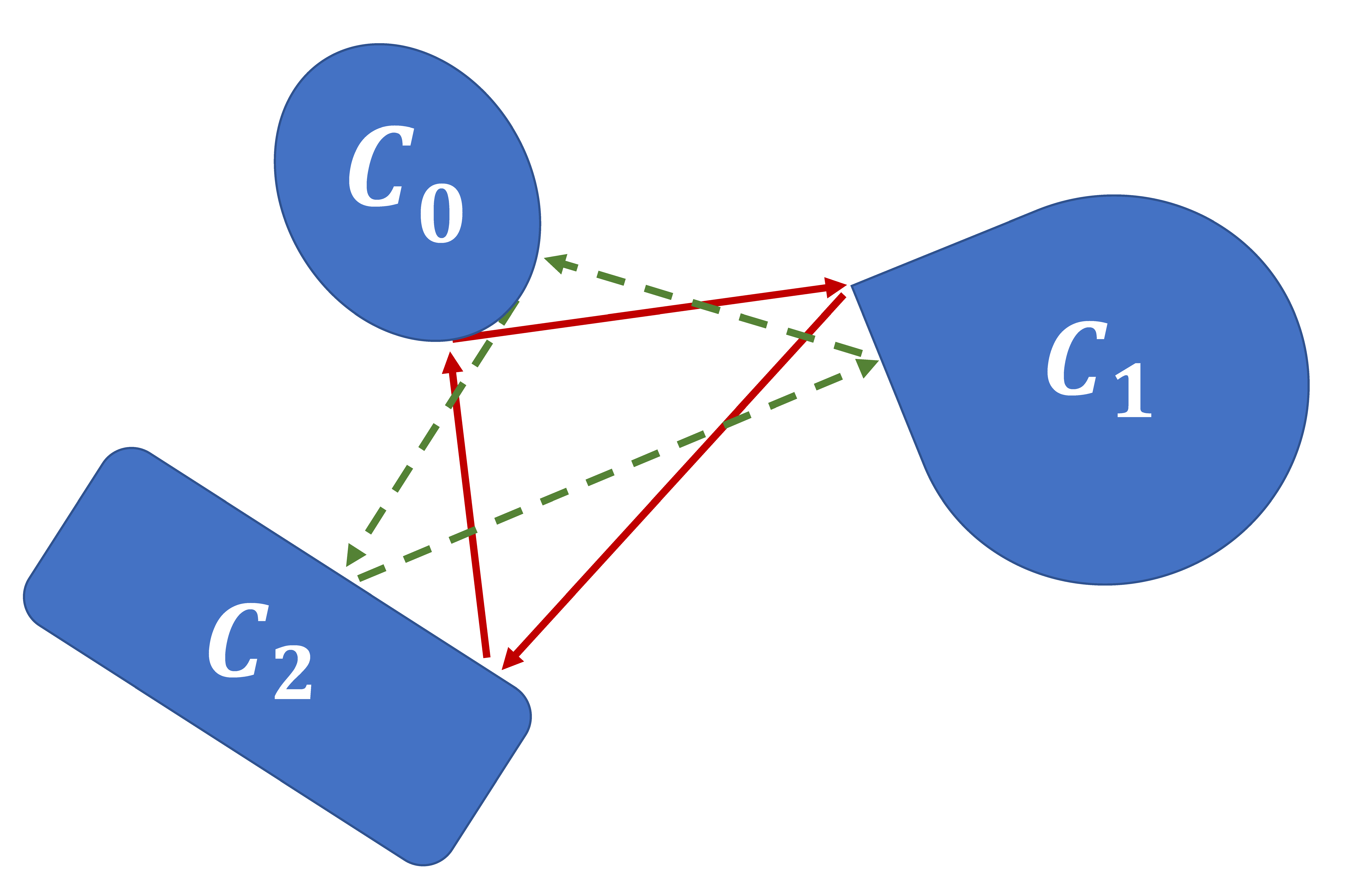}
		\caption{An illustration of two greedy limit cycles in alternating projections for non-intersecting convex sets. The limit cycles vary in accordance to the projection order. Here, projecting $C_0 \rightarrow C_1 \rightarrow C_2$ yields a different limit cycle to $C_2 \rightarrow C_1 \rightarrow C_0$. A simultaneous weighted projections approach is often used to resolve this issue and produce a unique result. Dilated POCS is an alternative approach.} 
		\label{fig_limit_cycles}
	\end{figure}

	When convex sets intersect, POCS can be visualized geometrically as in Figure~\ref{fig_MAP}.  Starting at any initialization, alternating projections will converge to a fixed point within the intersection of the sets. The intersection is also convex. If there are $I$ convex sets, then with $q$ as the iteration count the recursion,
	$$ \vec{x} [q+1] = P_{C_1} P_{C_2} \cdots P_{C_I} \; \vec x [q]$$
	will therefore converge to a point in the intersection of the sets
	$$ \vec{x}  [\infty] \in C_1 \cap C_2 \cap \cdots \cap C_I .$$

	When two convex sets do not intersect, POCS converges to a greedy limit cycle between the points in each set closest to the other set \cite{goldburg,youla1982}.
	
	When three or more convex sets do not intersect then no common solution exists and the alternating projections can approach a greedy limit cycle as seen in Figure~\ref{fig_limit_cycles}. The limit cycle can depend on the order of the projections. 
	
\begin{figure}[!t]
		\centering
		\includegraphics[width=2.5in]{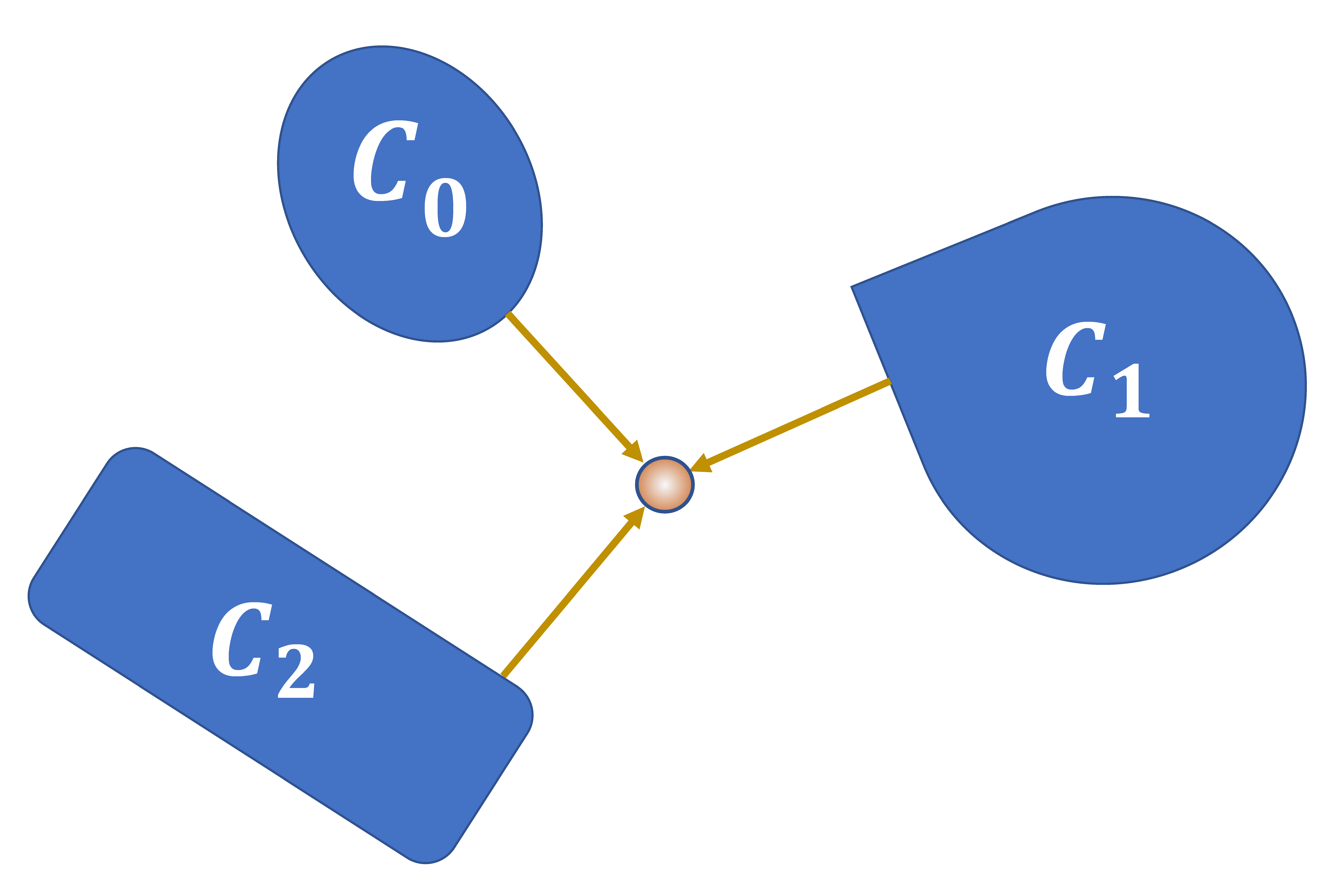}
		\caption{An illustration of the solution for a simultaneous weighted projection implementation of POCS when the sets do not intersect.}
		\label{fig_MAPx}
		\vspace{5mm}
		\includegraphics[width=3.3in]{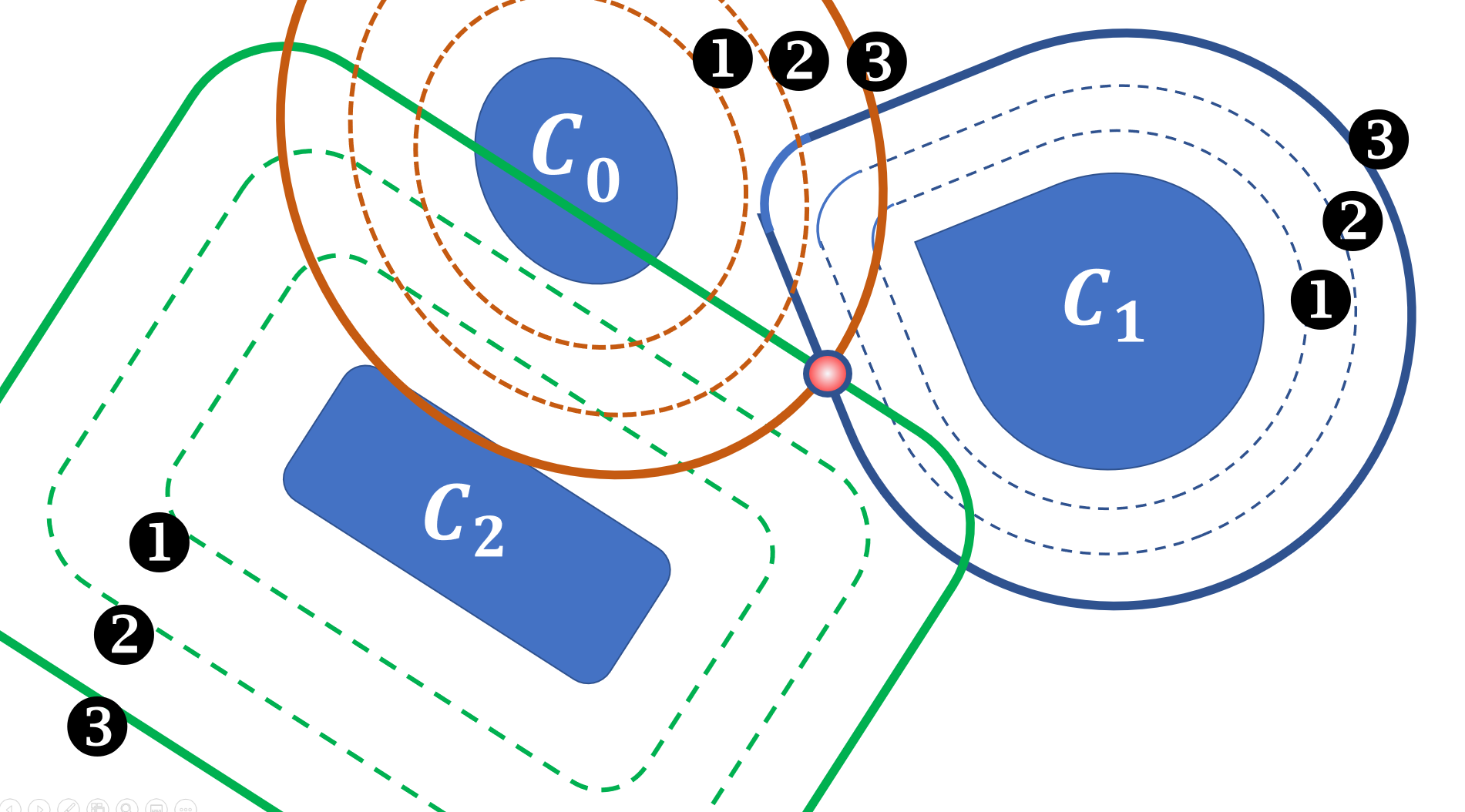}
		\caption{Nonintersecting convex sets can be dilated. The first dilation of the three sets, labled ``1'' in the figure, do not intersect. With further dilation to the dashed contours labeled ``2'', sets $C_0$ and $C_2$ intersect.  For the third dilation outlined by the solid lines, the three convex sets intersect at a single point.}
		\label{fig-Dilation}
	\end{figure}

	One solution to the nonintersecting sets uses simultaneous weighted projections. The importance of each set is determined by a weight $w_i$ such that $\sum_{i=1}^I w_i =1$ where $I$ is the number of the convex sets.   The iteration becomes
	$$ \vec{x} [q+1] = \sum_{i=1}^I w_i P_{C_i} (\vec{x} [q]).$$
	All projections are performed from the same point $\vec{x} [n]$ and then averaged using the weights. Convergence is to the MMSE solution
	$$ \vec{x} [\infty] = \arg \min_{\vec{y}}  \sum_{i=1}^I w_i \| \vec{y}- P_i (\vec{y})  \|.$$
	Such a solution is illustrated in Figure~\ref{fig_MAPx}.

	MMSE solutions give a compromise solution from conflicting design constraints. The minimax solution minimizes the worst case scenario. Both are viable optimization criteria based on signal synthesis roles.
	
MMSE solutions in POCS can generate unexpected and possibly undesired results. Consider the degenerate case of a single point being a convex set.
	Visualize twenty-one points corresponding to 21 simple convex sets where twenty of the points are clustered close together and the remaining point a long distance away. In an equally weighted simultaneous POCS MMSE solution, the outlier point will  have negligible effect on the solution and will essentially be ignored. The minimax solution obtained by dilated POCS, on the other hand, minimizes the maximum error and will lie midway between the cluster and the outlier. Which approach is best is  dependent on the problem and desired characteristics of the outcome.

	\section{Morphologically Dilating Signal Sets} 
	
	A brief overview of morphological dilation is appropriate to establish notation and provide background. The dual operation of erosion is discussed later in Section~\ref{EERode}.
	Tutorials are available in Giardina \& Dougherty \cite{Giardina} and Marks \cite{marks2009} (Chapter 12) 
	
\begin{figure}
		\centering
		\includegraphics[width=3.5in]{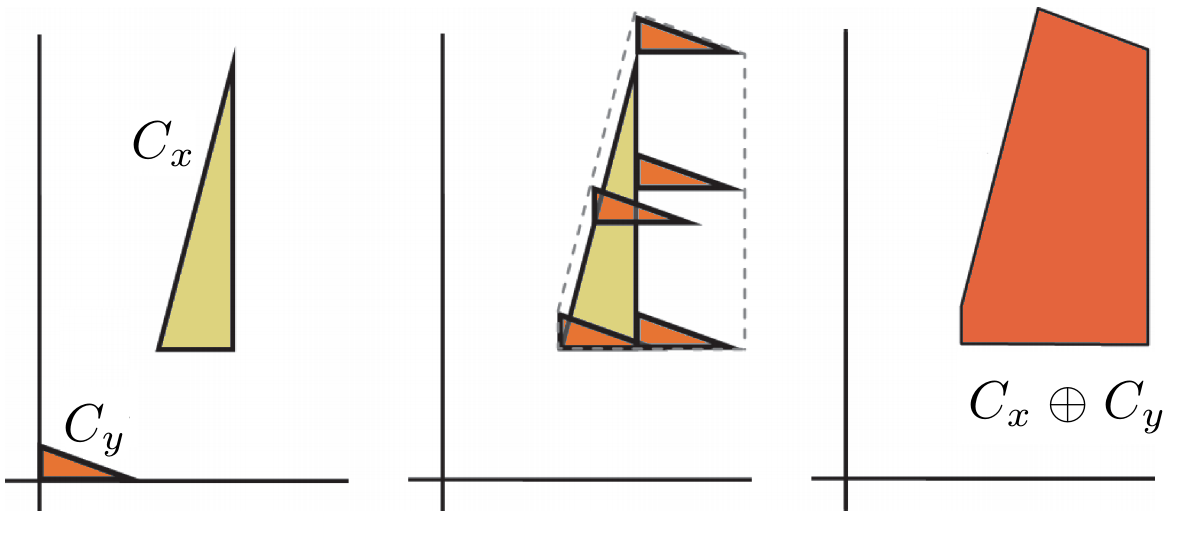}
		\caption{Dilation of the sets $C_x$ and $C_y$.}
		\label{Dilation}
	\end{figure}

	The dilation of sets $C_x$ and $C_y$, denoted as $C_x \oplus C_y$, is the set containing the elements of $\vec{x}+\vec {y}$ for all $\vec{x}\in C_x$ and  $\vec{y}\in C_y$ 
	\begin{equation}
		C_x \oplus C_y = \{ \vec{x}+\vec {y} \ | \ \vec{x} \in C_x, \vec {y} \in C_y \}.
		\label{dilute}
	\end{equation}
	Dilation is illustrated in two dimensions in Figure~\ref{Dilation}. If both $C_x$ and $C_y$ are convex, then so is the dilation set \cite{oh1993}. The set $C_y$ is referred to as the {\em dilation kernel}. 
	
	Morphological operations are most popularly applied to individual signals and images rather than to convex sets of signals as is done in dilated POCS.   
	
\begin{figure}%
		\includegraphics[width=3.5in]{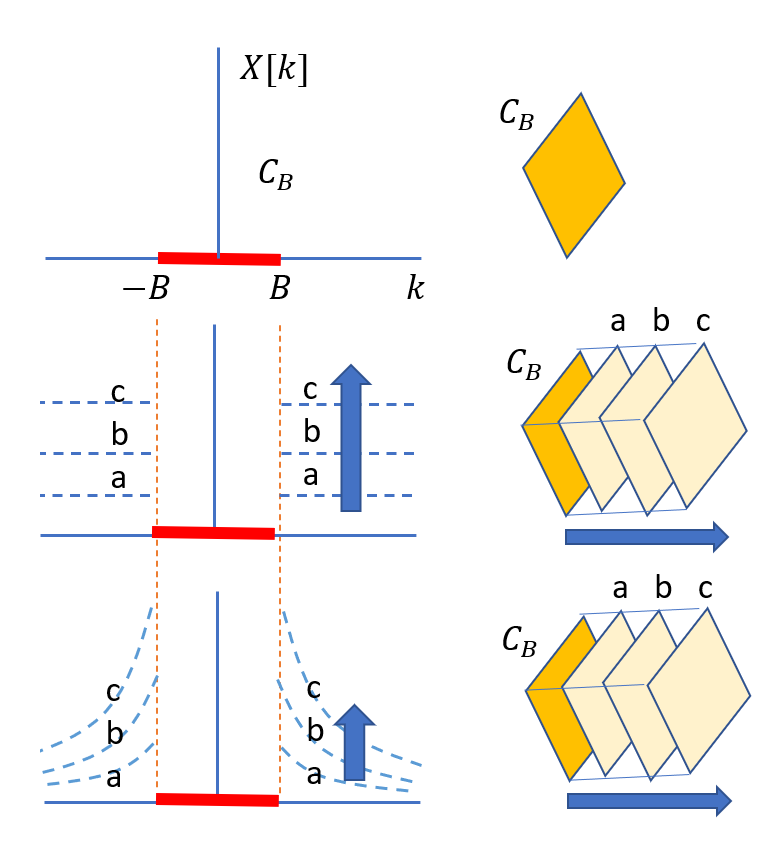}
		\caption{Signal space dilation example. As the bounds increase, the slab gets thicker.}
		\label{b}
	\end{figure}

	As an example of dilation of a set of signals, consider the convex set of bandlimited signals with a given bandwidth. Define the discrete Fourier transform (DFT) as\footnote{To ease visualization, the interval for $k$ is used to place the origin of the frequency domain, $k=0$, at the origin. This example thus requires $N$ to be odd.}  
	$$ X[k] = \sum_{n=0}^{N-1} x[n] e^{-j2\pi nk/N}; \;
	|k|\leq \frac{N-1}{2}.$$	
	Let $C_B$ denote the set of all signals where $X[k]=0$ for $|k| > B$. This is the set of bandlimited signals with bandwidth $B$. Thus 
	$$C_B =\left\{ x[n] \ | \ X[k]=0 \mbox{ for } |k|> B
	\right\}.$$
	As illustrated in the top row of Figure~\ref{b}, the set $C_B$ is a (convex) plane passing through the origin \cite{marks2009}. 
	
	Dilation of a convex set of signals can often be done in many ways. To illustrate, we now show three ways to dilate the convex set of bandlimited signals, $C_B$.
	
	\begin{enumerate}
	    \item 
	Define set $C_a$ as the set of all signals whose Fourier transform real part lies between zero and an upper bound $a$. Thus
	$$C_a =\left\{ x[n] | 0 \leq  \Re X[k]  \leq a 
	\right\}$$
	where $\Re$ denotes the real part operation. The set $C_a$ is a box in the first orthant with a vertex at the origin and with linear dimension $a$. Dilation of a plane with a box generates a slab, i.e, all the points between two parallel planes.\footnote{
		Dilation of a line, i.e., a one dimensional plane, is a cylinder with a cross-section dependent on the dilation kernel.}
	If $a$ is replaced by $b>a$, the slab becomes thicker. This is illustrated in the middle row in Figure~\ref{b}. Choosing $c>b$ results in an even thicker slab.
	
	\item The bottom row in Figure~\ref{b} shows tapered bounds can also be used for dilation resulting in a similar dilation in signal space. Since $b>a$, the dilated convex set $C_a$ is a subset of set $C_b$. In the fashion illustrated here, the ever increasing dilations in Figure~\ref{fig-Dilation} can be generated in general.  
	
	\item A third method of dilation of the set of bandlimited signals, not pictured, can be done by increasing the bandwidth from $B$ to $B_a>B$.  The dilation kernel set can be defined as
	$$ C_{B,B_a}=\left\{ x[n] \ | \ X[k]=0 \ \forall \  k \notin 
	B<|k| \leq B_a \right\}.$$
	Like $C_B$, the set  $C_{B,B_a}$ is a hyperplane. Visualize two lines in three space that pass through the origin. The dilation on one of these lines with the other results in a plane that includes both lines, i.e., the span of the two vectors. The dilation in general results is the hyperplane set    
	$$ C_{B_a} = C_B \oplus C_{B,B_a} =  \left\{ x[n] | X[k]=0 \mbox{ for } |k|> B_a
	\right\}.$$
	\end{enumerate}
	
	 These three examples illustrate there can be numerous ways to perform dilation on a given convex set of signals. In practice, domain expertise can be applied to choose the method most effective for a given problem. 
	

	\section{Low Rank Matrix Solution}
	
	To illustrate MMSE versus minimax solutions, consider the matrix equation 
	\begin{equation} 
		\vec{y}= {\bf A} \vec{x}
		\label{matrix}
	\end{equation}
	where the matrix $\bf{A} $ has more rows, $L$, than columns, $K$. Consider a vector $ \vec{y}$  for which there exists no vector $\vec{x}$ that satisfies (\ref{matrix}). Finding a vector $\vec{y}$ in some sense close to the solution of (\ref{matrix}) is an inverse problem. The definition of ``close'' can be defined either by the MMSE or the minimax metrics.

	\subsection{The MMSE Solution}
	For any $\vec{x}$, the value of $\vec{y}$ must live on a plane defined by the column space of $\bf{A}$. To show this, divide the matrix $ {\bf A}$ into columns.
	$$ {\bf A}=\left[ \vec{a}_1 \;\vec{a}_2   \ldots \vec{a}_k \ldots    \vec{a}_K  \right].$$
	Then
	$$ \vec{y} =  \sum_{k=1}^K \vec{a}_k \; x_k .$$
	The vector $\vec{y}$ must then always be composed of weighted combinations of the column vectors $\vec{a}_k$. The span (the set of all linear combinations) of the column vectors form a plane, or subspace, in $L$ space. This is the matrix's  column space. It is a $K$ dimensional plane in an $L>K$ space and is defined as
	$$ C_{\bf A}  = \left\{ \vec{y} \; | \; \vec{y}={\bf A} \vec{x} \ \forall \vec{x}  \right\}. $$ 
	For a vector $\vec{y}$ not lying in the space, the best solution for $\vec{x}$ that satisfies (\ref{matrix}) is the orthogonal projection of $\vec{y}$ onto the column space $C_{\bf A}$. The optimal value of $\vec{x}$ is the solution to
	
	\begin{equation} \vec{x}_{\mbox{MMSE}}= \arg \min_{\vec{x}} \| \vec{y}- {\bf A} \vec{x} \| .
		\label{MSE}
	\end{equation}
	Assuming the rank of the matrix is the same as the number of columns (i.e., ${\bf A}^T {\bf A}$ is nonsingular), the Moore-Penrose inverse solution \cite{Tewarson} is
	
	\begin{equation} \vec{x}_{\mbox{MMSE}}= ({\bf A}^T {\bf A})^{-1} {\bf A}^T \vec{y}.
		\label{SuddoInv}
	\end{equation}
	This is the MMSE solution.
	
	\subsection{MiniMax}
	
	The minimax (mM) solution of (\ref{matrix}) is  
	\begin{equation}
		\vec{x}_{\mbox{mM}}= \arg \min_{\vec{x}}  \max_k | y_k - \vec{a}_k^T \vec{x} |.
		\label{mmGood}
	\end{equation}
	The minimax solution can be obtained using dilated POCS. 
	For the corresponding minimax solution, let $\vec{r}_\ell$ denote the $\ell$th row of ${\bf A}$. Note that $\vec{r}_\ell$ is a row vector. The matrix relationship in (\ref{matrix}) can then be expressed as the inner products
	\begin{equation} y_\ell =  \vec{r}_\ell \vec{x}; \ 1\leq \ell \leq L.
		\label{MM11}
	\end{equation}
	These are $L$ affine planes \cite{Piziak} (a.k.a. linear varieties \cite{Luenberger}) in $\vec{x}$ space.   
	
	The  $\ell$th dilated convex set is
	\begin{equation}  y_\ell -\varepsilon \leq \vec{r}_\ell \vec{x} \leq  y_\ell +\varepsilon  
		\label{interval}
	\end{equation}
	where $\varepsilon$ denotes a scalar dilation parameter. Equivalently,
	$$ \left| \vec{r}_\ell \vec{x} - y_\ell \right| \leq \varepsilon.$$
	Dilated POCS seeks to adjust $\varepsilon$ to the smallest value satisfying this equation for all $\ell$.
	
	The projection of a vector $\vec{w}$ onto the $\ell$th linear variety in \eqref{MM11} is \cite{marks2009} (page 510) follows as 
	$$ P_\ell \vec{w} =   \vec{w} - \frac{\vec{r}_\ell^T}{\| \vec{r}_\ell^T \|^2}  
	\left( \vec{r}_\ell \vec{w} -y_\ell \right).
	$$ 
	The projection onto the $\ell$th dilated linear manifold follows as 
	$$ P_{D\ell} \vec{w} =  
	\left\{
	\begin{tabular}{ll} 
		$ \vec{w} - \frac{\vec{r}_\ell^T}{\| \vec{r}_\ell^T \|^2}  
		\left( \vec{r}_\ell \vec{w} - (y_\ell - \varepsilon) \right) $		& ;   $\vec{r}_\ell \vec{w} -y_\ell <  - \varepsilon $ \\
		$ \vec{w}
		$			
		& ; $ \left| \vec{r}_\ell \vec{w} - y_\ell \right| \leq \varepsilon$ \\
		$ \vec{w} - \frac{\vec{r}_\ell^T}{\| \vec{r}_\ell^T \|^2}  
		\left( \vec{r}_\ell \vec{w} - (y_\ell + \varepsilon) \right) $		& ; $\vec{r}_\ell \vec{w}-y_\ell  > 
		\varepsilon.$ \\
	\end{tabular}
	\right.	
	$$

	\subsection{Matrix Example}\label{MatExSec}
	
	As an example, consider the matrix
	
	$$ {\bf A} = \left[ 
	\begin{tabular}{ccccc}
		1 &  0 & 1 &1 &1 \\
		0 &1 & 1 & 1  &1 
	\end{tabular}
	\right]^T
	$$
	and
	$$ \vec{y} = \left[ \  0 \ 0 \ 1 \ 2 \ 7 \ \right]^T$$
	where the superscript $T$ denotes matrix transposition. There is no vector $\vec{x}$ that satisfies $ \vec{y} = {\bf A} \vec{x}$.
	
	\subsubsection{MMSE} From (\ref{SuddoInv}) the MMSE solution of (\ref{MSE}) is
	\begin{equation} \vec{x}_{\mbox{MMSE}} = ({\bf A}^T {\bf A})^{-1} {\bf A}^T \vec{y} = \left[
		\begin{tabular}{c}
			1.4286 \\ 1.4286 
		\end{tabular}
		\right].
		\label{sudo2}
	\end{equation}
	The solution is shown in Figure~\ref{MatrixExample}. The corresponding error is
	$$ \| \vec{y} - {\bf A} \vec{x}_{\mbox{MMSE}} \| = 5.0427.$$
	
	\subsubsection{Minimax} 
	
\begin{figure}[!t]
		\subfloat[Predilation example sets. The minimax and minimum MMSE is shown.   ]{
			\includegraphics[width=3.2in]{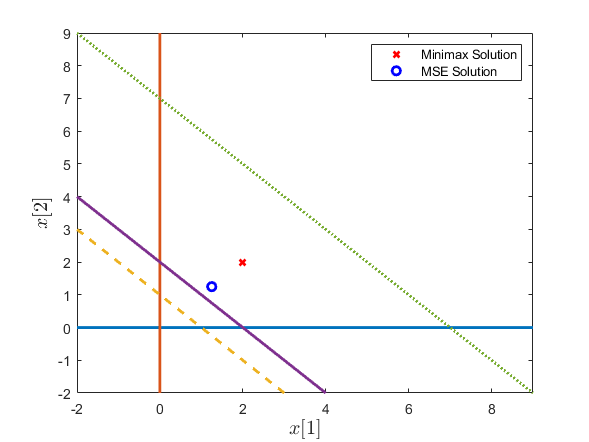}%
			\label{MatrixExample1}}
		\vfill
		\subfloat[Example sets with the most constraining dilations shown]{
			\includegraphics[width=3.2in]{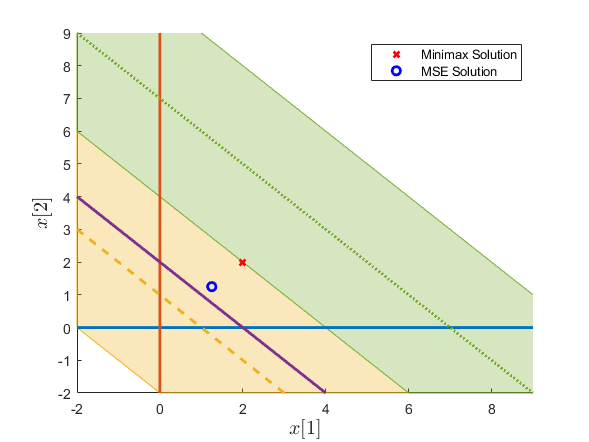}%
			\label{MatrixExample2}}
		\caption{Visualization of the constraints from section \ref{MatExSec}. The sets from the $3^{rd}$ and $5^{th}$ rows of {\bf A} are the most distant and shown as a yellow dashed and a green doted lines respectively. The dilated slabs for these two sets are also shown in (b) ($\varepsilon = 3$), note that the minimax solution lies on the line of their intersection.}
		\label{MatrixExample}
	\end{figure}

	The minimax solution in (\ref{mmGood}) as illustrated in Figure~\ref{MatrixExample} is
	\begin{equation}
		\vec{x}_{\mbox{mM}} = 
		\left[
		\begin{tabular}{c}
			2 \\  2 
		\end{tabular}
		\right].
		\label{mmSol2AG}
	\end{equation} 
	
	To see this,  five lines in (\ref{MM11}) are plotted in Figure~\ref{MatrixExample} corresponding to the five equations - one for each row of $\bf A$. Each line is dilated into a slab until intersection, in this case when  $\varepsilon = 3.0$. The two lines that require the most dilation for intersection, rows three and seven of {\bf{A}}, are shown in Figure~\ref{MatrixExample}.
	
	Dilation occurs until all three sets contain the common point given in (\ref{mmSol2AG}). Both $\vec{x}_{\mbox{mM}}$ and $\vec{x}_{\mbox{MMSE}}$ lie inside the largest triangle in Figure~\ref{MatrixExample}. Because two of the constraint lie on the axes and the rest are parallel to each other, both also lie on the $45^o$ line of the first quadrant.

	
	\section{Computed Tomography}
	
\begin{figure}[!t]
		\centering
		\includegraphics[width=3.2in]{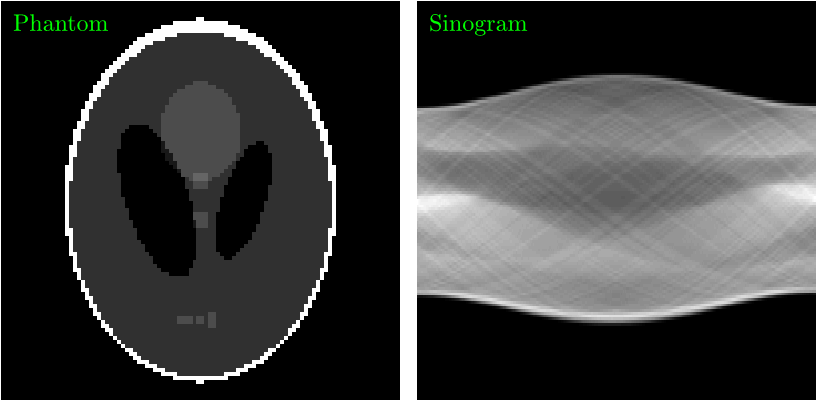}
		\caption{A $100\times 100$ modified Shepp-Logan phantom and its corresponding sinogram, sampled at 180 one-degree angles. The Shepp-Logan phantom is a popular test image for image construction representing the features of the human brain and skull.  Here, the phantom used is a normalized greyscale image within [0,1].}
		\label{fig_phantom_true}
	\end{figure}

	We apply dilated kernels to the issue of medical imaging via computed tomography (CT) reconstruction. Here, a target object is imaged at various angles resulting in projection slices corresponding to the time-delay of the scan passing through the imaged body from specific viewpoints. These projections from various angles form the object’s profile or sinogram as shown in Figure~\ref{fig_phantom_true}, where the target object used is a $100\times 100$ image of the modified Shepp-Logan phantom. A beam path matrix is used to model this interaction \cite{Penfold2009}. 
	However, the problem becomes large-scale with increasing pixel resolution, necessitating efficient, parallelizable inversion techniques.  POCS-based solutions include the algebraic reconstruction technique (ART) \cite{gordon1970,marks2009,Penfold2010}. The simultaneous iterative reconstruction technique (SIRT) and simultaneous algebraic reconstruction technique (SART) \cite{anders1984} are based on alternating and averaged projections. Alternatively, filtered back-projection (FBP) \cite{marks2009} reconstructs by back-projecting the sinogram under a ramp filter.  Due to issues with noise and streak artifacts, FBP is generally replaced with iterative methods in practice.  

	For an $m\times n$ path matrix, $n$ corresponds directly with the number of pixels in the image, while $m$ is the product of the length of projection bins and the number of angles tested. For the simple $100\times 100$ phantom in Figure~\ref{fig_phantom_true} using 180 one-degree slices into 1-pixel wide bins results in a sparse matrix with dimensions $26100\times 10000$.  In the following examples SART is used to form the least-squares solution while dilated POCS is used for the minimax solution.  Image recovery error is measured with respect to the given sinogram and the reconstructed image's sinogram, not the reconstructed image itself.

	\begin{figure}[!t]
		\centering
		\includegraphics[width=2.8in]{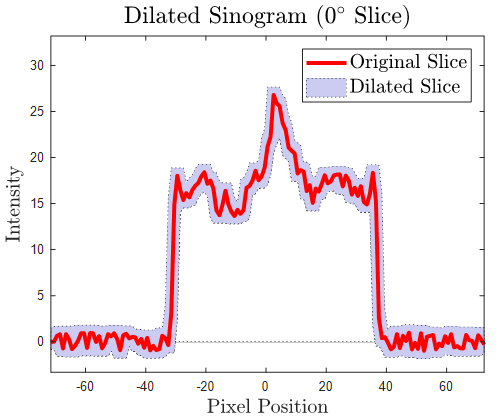}
		\caption{The $0^\circ$ slice of the sinogram, dilated with a box corresponding to lateral vibration and signal noise. The noise dilation changes with each iteration as the solver ascertains the noise level while maximum lateral movement is constrained. Rather than project onto the given (red) sinogram itself, the current iteration of the reconstructed image is projected to within the boundaries of the dilated slice instead.}
		\label{fig_phantom_slice}
	\end{figure}
	\begin{figure*}[!t]
		\centering
		\subfloat[Original (Unknown) and Noisy Sinogram (Given)]{
			\includegraphics[width=3.2in]{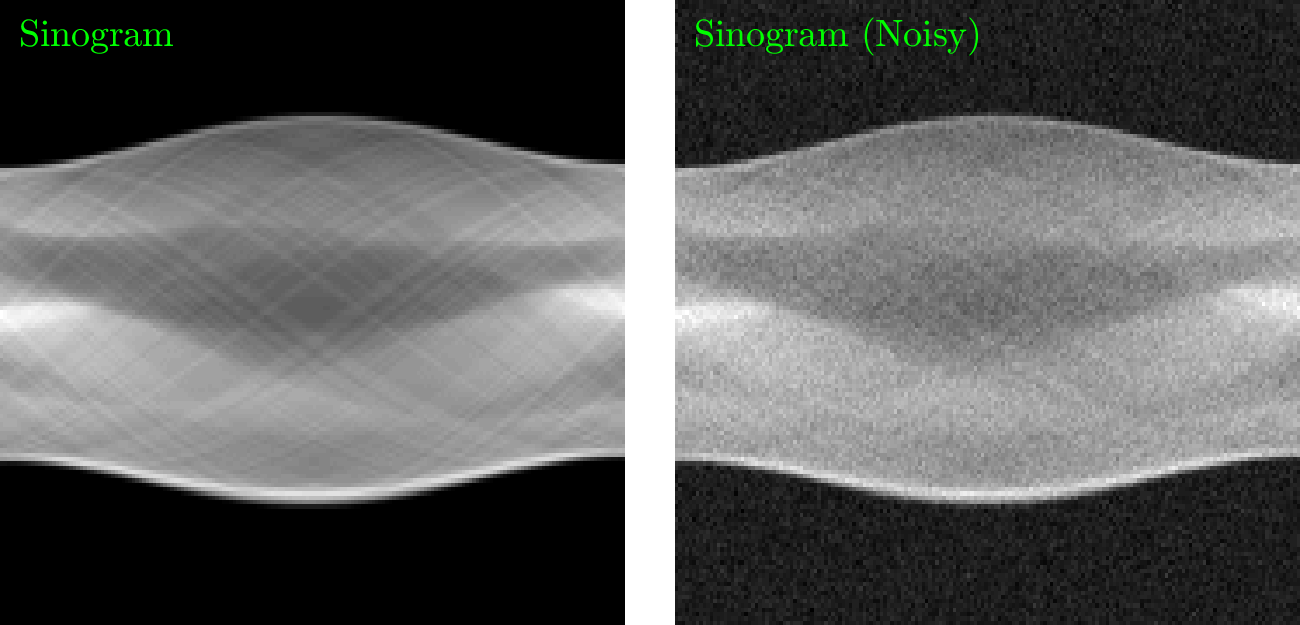}%
			\label{fig_gaussian_sino}}
		\hfill
		\subfloat[Dilated Recovery Image and Sinogram]{
			\includegraphics[width=3.2in]{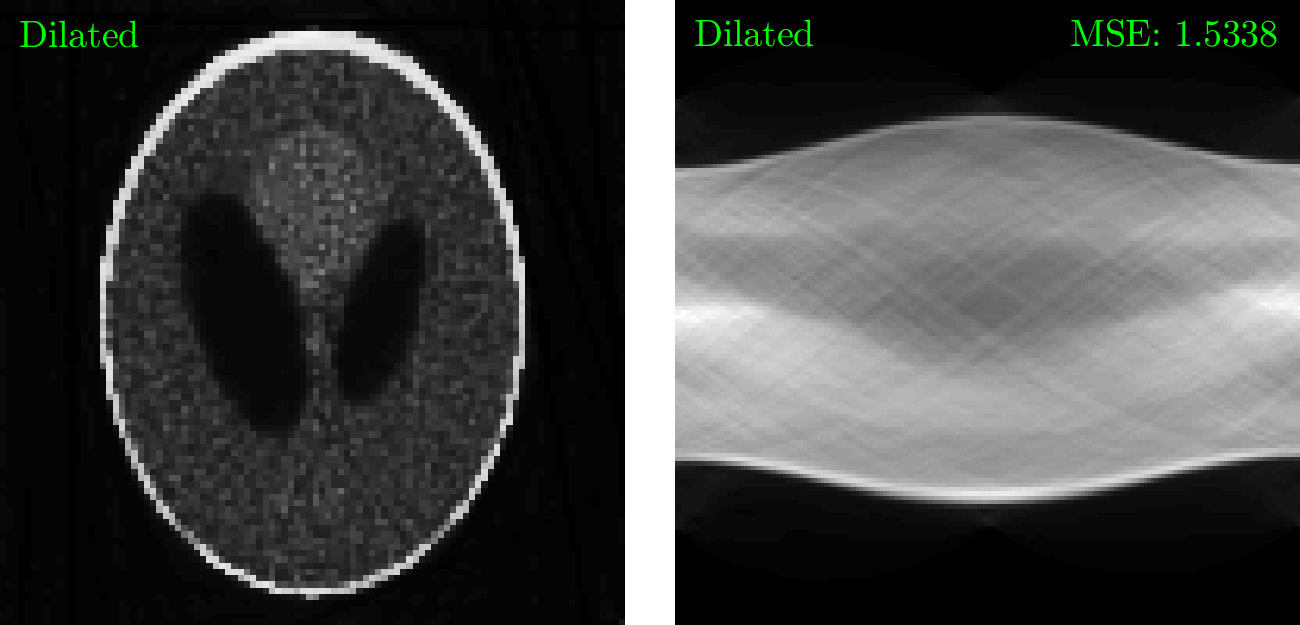}%
			\label{fig_gaussian_DIL}}
		\vfill
		\subfloat[MMSE Recovery Image and Sinogram]{
			\includegraphics[width=3.2in]{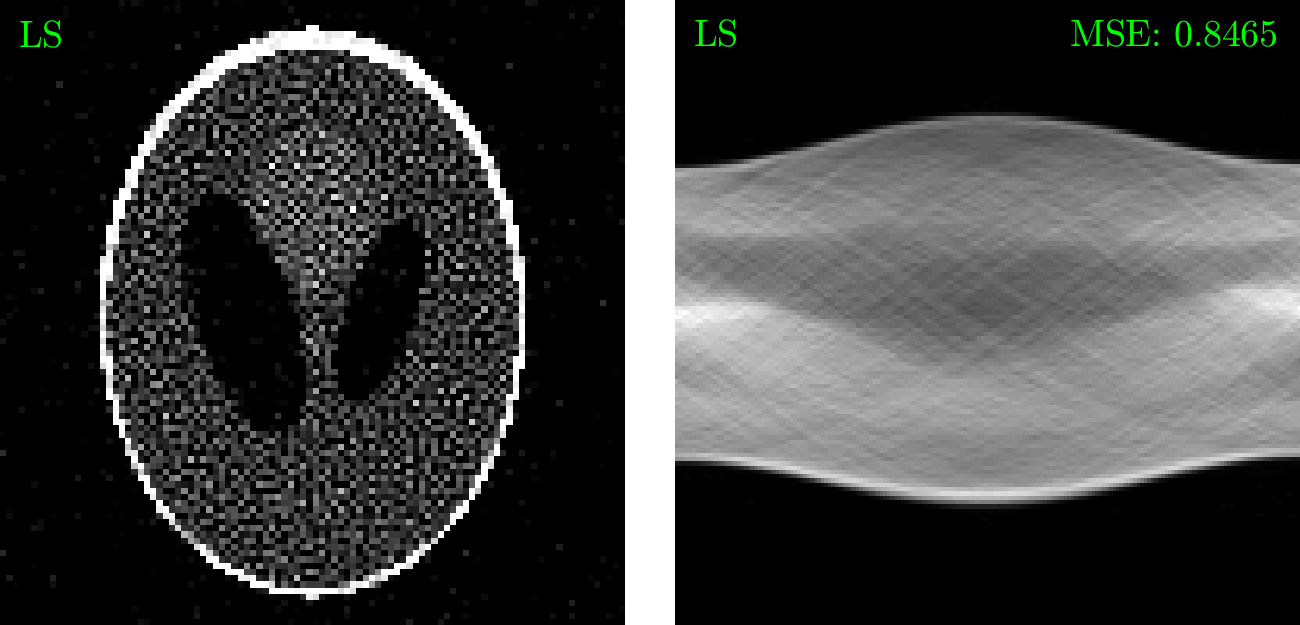}%
			\label{fig_gaussian_LS}}
		\hfill
		\subfloat[Filtered Back-Projection Image and Sinogram]{
			\includegraphics[width=3.2in]{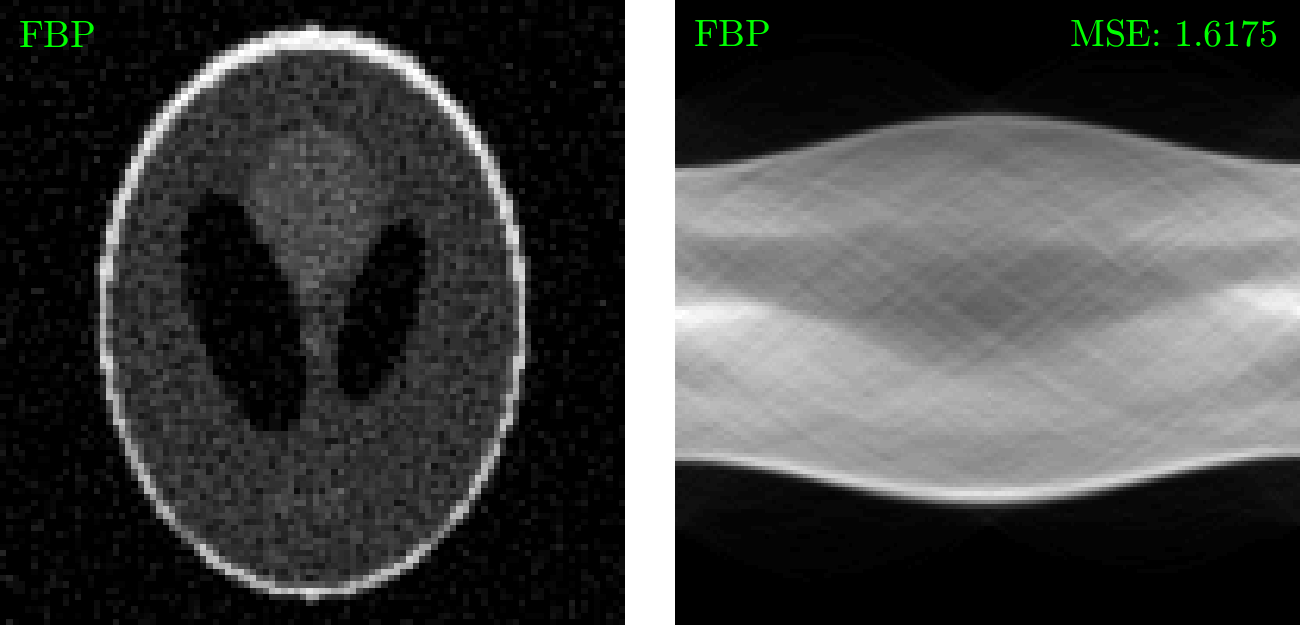}%
			\label{fig_gaussian_fbp}}
		\caption{The Shepp-Logan phantom sinogram (a) is corrupted by Gaussian random noise in with $\sigma=1.0$ and is used to reconstruct an image using (b) dilation, (c) MMSE algorithms, and (d) filtered back-projection. The MMSE sinogram is the closest $L^2$-norm to the given noisy sinogram, but the dilated result displays cleaner edges and features.}
		\label{fig_gaussian}
	\end{figure*}
	\subsection{Box-Dilated Sinogram}
	
	This CT example demonstrates how box-dilation is situationally advantageous to ball kernels.  Rather than truncate the projected step, the sinogram constraint may be dilated instead.  The vectorized sinogram can be dilated horizontally for lateral movement and path errors of the imaged object, or vertically for noise. An exclusively vertical dilation of the sinogram is the classic minimax error problem \cite{Karbasi}. A slice of this noisy sinogram for the Shepp-Logan phantom reconstruction is seen in Figure~\ref{fig_phantom_slice}. This box-dilation can be reshaped to emphasize noise or lateral movement by removing slack in the other dimension.
	
	\subsection{CT Reconstruction Results}
	
\begin{figure*}[!t]
		\centering
		\subfloat[Original (Unknown) and Noisy Sinogram (Given)]{
			\includegraphics[width=3.2in]{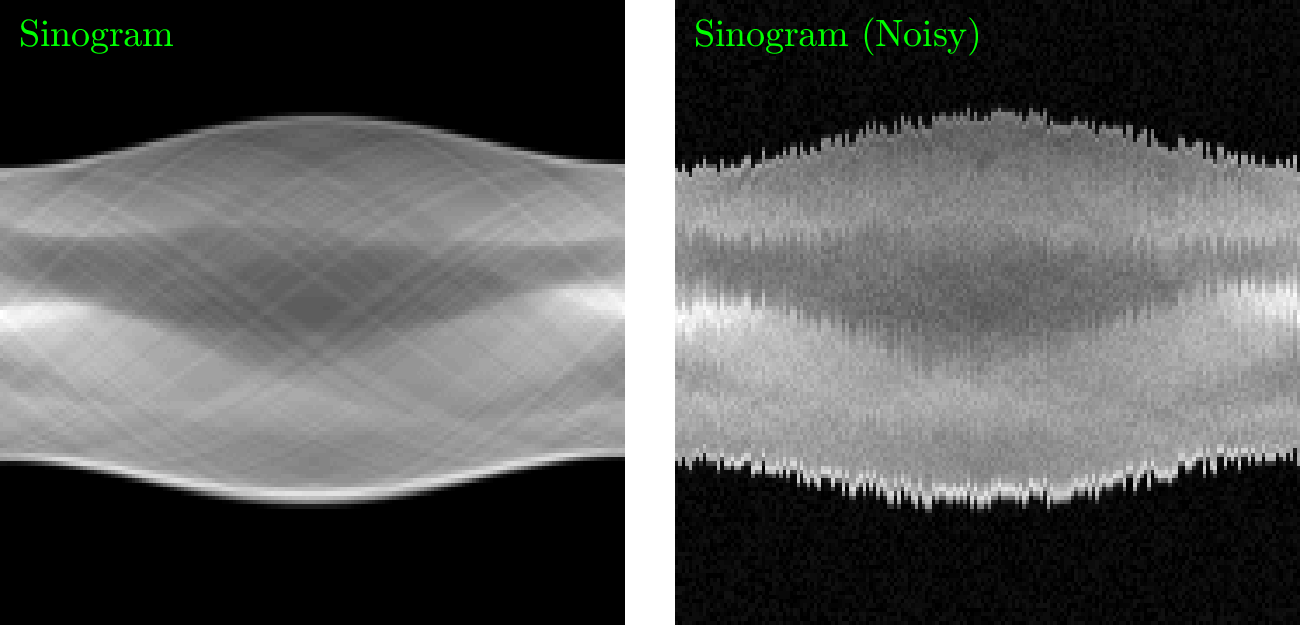}%
			\label{fig_wob_noisy}}
		\hfill
		\subfloat[Dilated Recovery Image and Sinogram]{
			\includegraphics[width=3.2in]{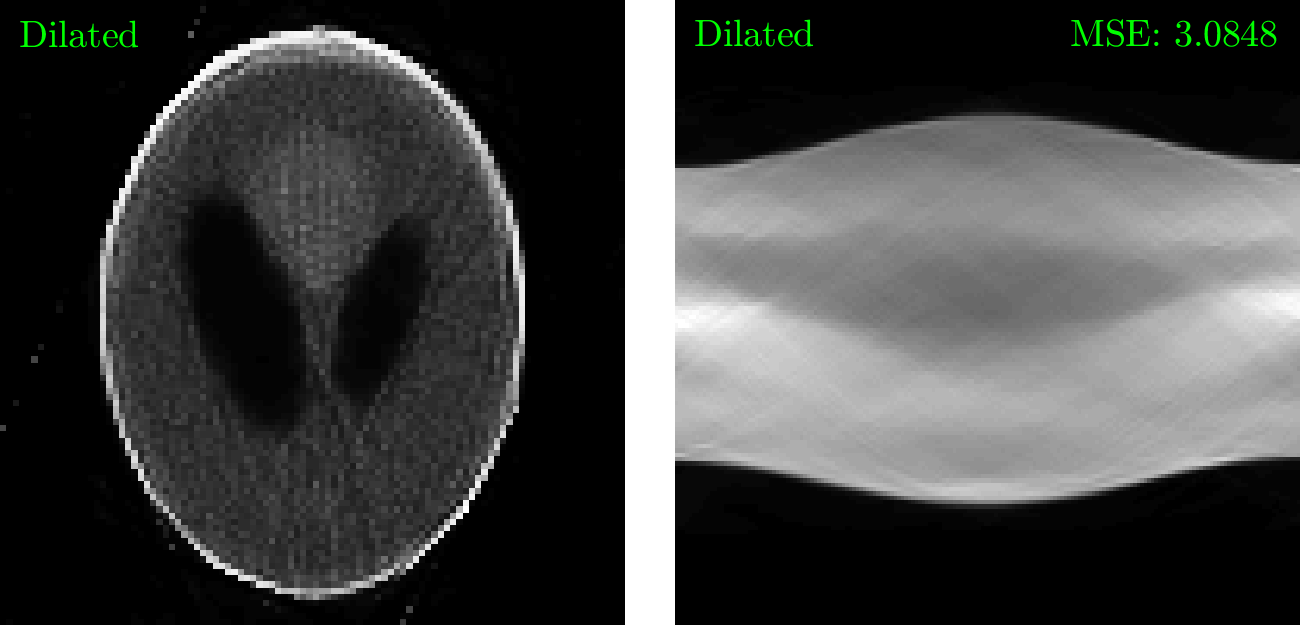}%
			\label{fig_wob_dil}}
		\vfill
		\subfloat[MMSE Recovery Image and Sinogram]{
			\includegraphics[width=3.2in]{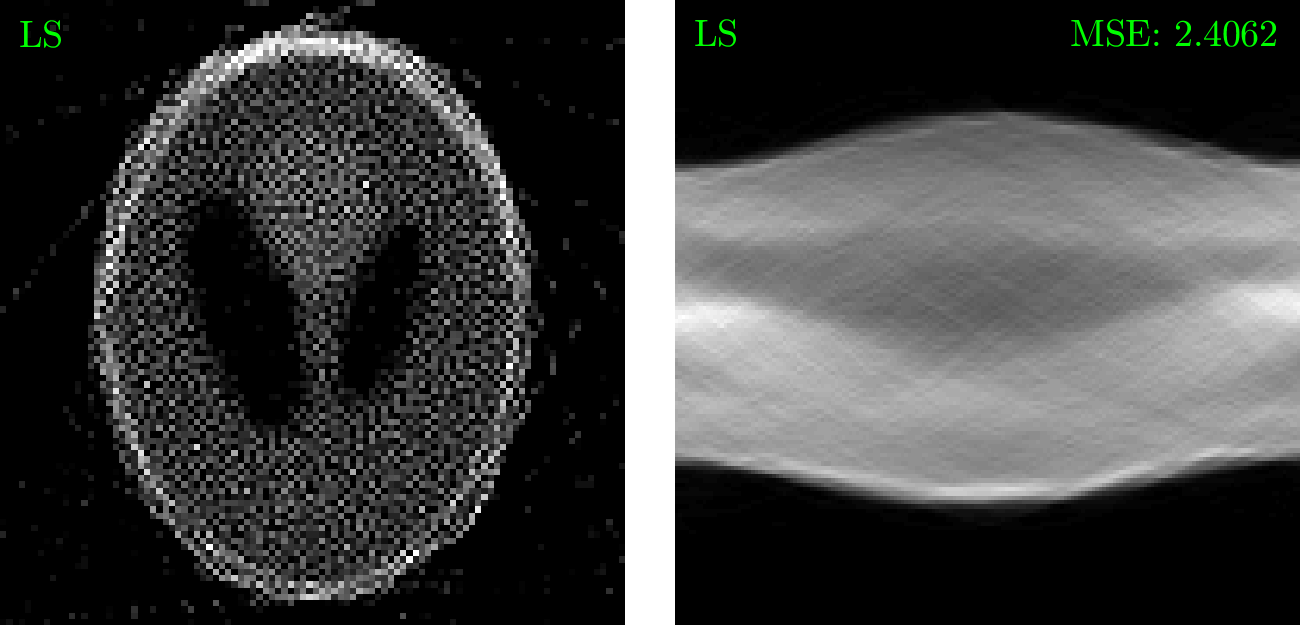}%
			\label{fig_wob_ls}}
		\hfill
		\subfloat[Filtered Back-Projection Image and Sinogram]{
			\includegraphics[width=3.2in]{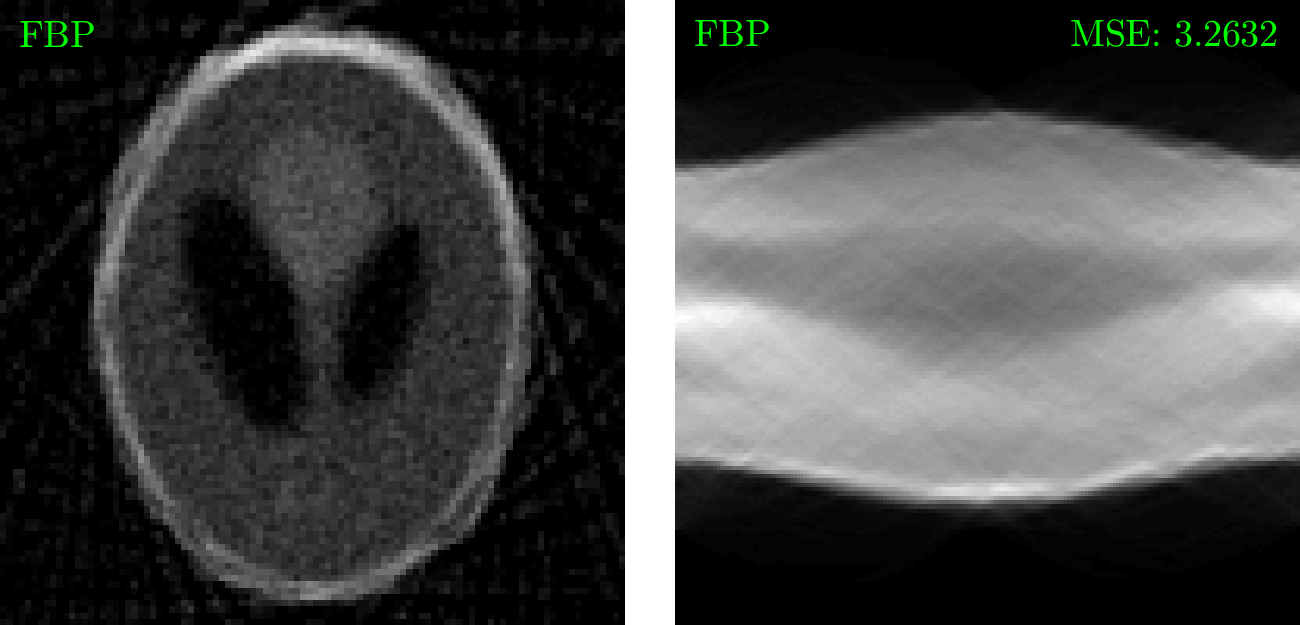}%
			\label{fig_wob_fbp}}
		\caption{The Shepp-Logan phantom sinogram (a) is corrupted by uniform random noise in [-1,1] and uniform random lateral motion in [-2,2] pixels.  Reconstructions are made using (b) dilation, (c) MMSE algorithms, and (d) filtered back-projection. Although by design the least-squares sinogram has the lowest mean square error, but the dilated result's image has the cleanest features. Streak artifacts from the target's motion are seen in the MMSE and FBP reconstruction.}
		\label{fig_wobble}
	\end{figure*}

	The results of the dilated algorithm is compared with a MMSE technique and FBP.  The recovered images and sinograms are displayed in Figures~\ref{fig_gaussian} and \ref{fig_wobble}. Figure~\ref{fig_gaussian} demonstrates the slight advantage of dilated kernels for zero-mean Gaussian noise. The MMSE algorithm is applied to the reconstructed image's sinogram to the observed sinogram, but this does not translate to coherent recovered images. Projecting onto the dilated sets yielded advantageous results.  Figure~\ref{fig_wobble} introduces lateral shifts in addition to zero-mean uniform noise to the sinograms, resulting in the application of box kernels for dilation. For the Shepp-Logan phantom, the dilated result produced a grainy image with sharp edges and suppressed streak artifacts.

	\section{Signal Set Erosion}
	\label{EERode}
	
	Dilation makes sets bigger. Erosion makes them smaller.
	
	Erosion can be defined using complementary sets. Let $\overline{C}$ be the complement of $C$. In other words, $\overline{C}$ consists of all $\vec{x}$ not in $C$. Dilating the complement of $C_x$ with $C_y$ and taking the complement of the result gives the erosion of $C_y$ by $C_x$. In other words
	$$ C_y \ominus C_x = \overline{ \overline{C_x} \oplus C_y } .   $$
	
\begin{figure}%
		\begin{center}
			\includegraphics[width=2.5in]{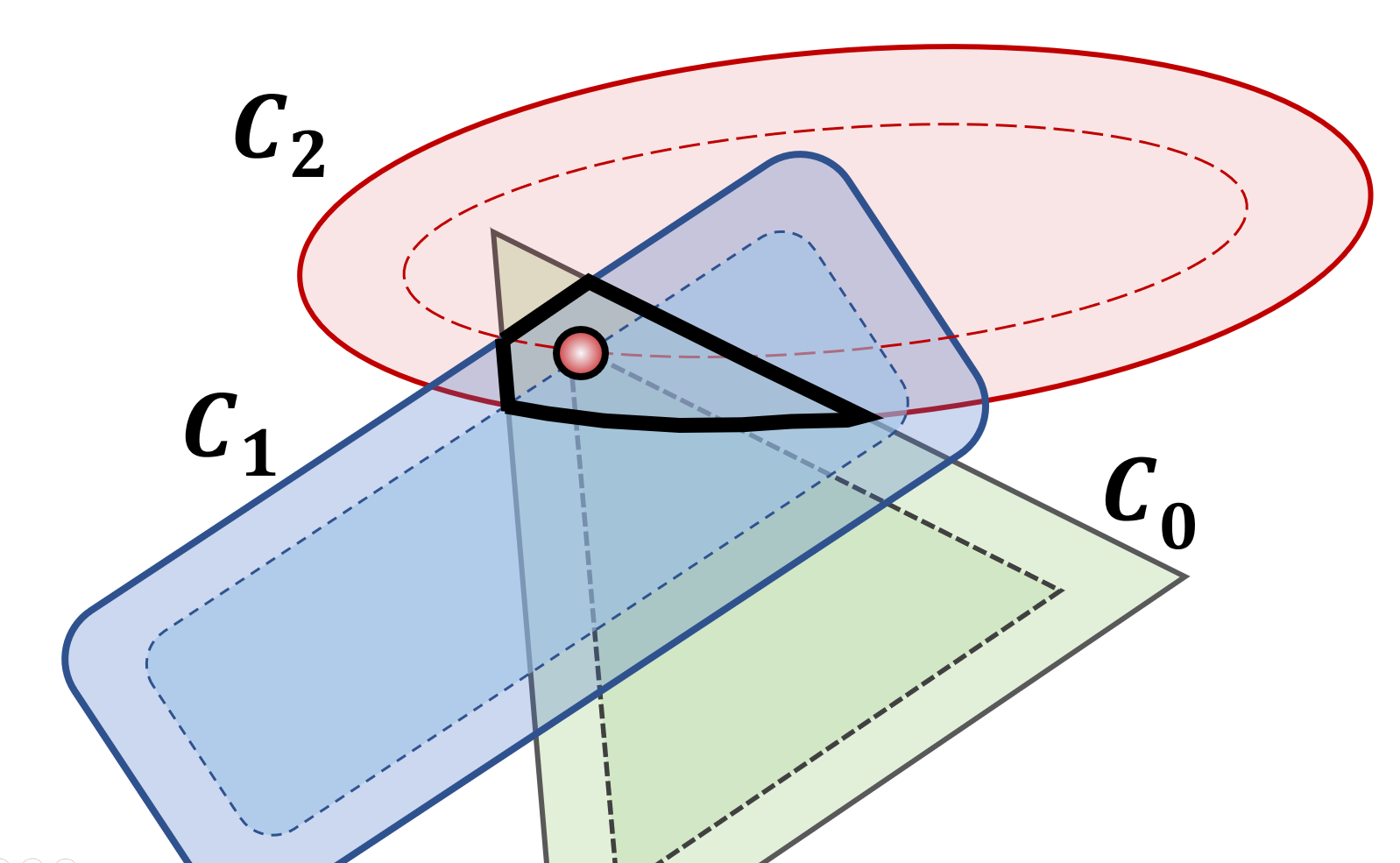}
			\caption{Illustration of erosion of convex signal sets. The thick bold line outlines the intersection of the three sets. Depending on initialization, POCS can converge to any point in this intersection. Here the three convex sets are eroded so that the three erosions intersect in at a single point. The erosions are outlined by dashed lines.  Conventional alternating projection among these eroded sets will converge to this single point independent of the POCS initialization.}
			\label{Erode}
		\end{center}
	\end{figure}

	
	If $C_x$ and $C_y$ are convex, the erosion $C_y \ominus C_x$ appears also to be convex. As with dilation, some sets of signals can be straightforwardly eroded. Figure~\ref{Erode} illustrates an advantage of doing so. Three convex sets intersect in the region outlined by the bold black line. Conventional POCS, applied to these signal constraints can converge to any point in this intersection depending on the iteration initialization. The set of solutions can be narrowed by eroding each of the signal sets as shown. When convex constraints intersect at a single point, conventional POCS will converge to the single point independent of initialization. The single point is obtained by appropriately eroding each set. The eroded sets are outlined by the dashed lines in Figure~\ref{Erode} that intersect in a single point.   
	
	
	\section{Dilation Optimization}
	
	Insufficient dilation in dilated POCS results in one or more of the sets not intersecting. Over dilation gives an overly large set intersection.
	
	The optimal dilation can be found using interval halving \cite{Hays2017,Hays2018}. Insufficient dilation can be identified by the POCS entering a limit cycle. For both the optimal and over dilated cases, POCS will converge to a single solution dependent on initialization. A solution between an under and over dilated case can be identified and a mid point dilation applied. Whether or not this mid point solution converges or cycles dictates whether it will replace the under or over dilated point in the next interval halving.
	
	Note that a proper dilation solution need not intersect at a single point. Visualize, for example, two parallel lines on a plane corresponding to two convex sets. Dilation of the two lines into slabs can give the smallest overlap when the intersection of dilations touch along a third parallel line. The best dilation solution is here a line rather than a point.  
	
	There are other cases where convex set dilations intersect only in the limit. An example is continuous time limited signals and bandlimited signals. Both define convex signal sets \cite{marks2009} and can be dilated by increasing the internal where the functions are nonzero. Famously, time limited signals cannot be bandlimited and visa versa \cite{Slepian1976}. Thus, a bandlimited signal can be time limited only when the time domain interval contains the entire real line. As bandwidth and the time limit interval are increased, the dilated convex sets will move ever closer together but never intersect except in the limit. 
	
	\section{Conclusion}
	
	Dilated projections can enhance the outcome of POCS in situations where multiple, non-intersecting convex constraint sets are dictated. Instead of converging to a MMSE error, dilated projections can achieve minimax weighted error solutions that can have applications in solving over-constrained systems of equations and tomographic image reconstruction. 
	
	Unlike simultaneous weighted projections, there are no explicit weights in dilated POCS. Weighting of the importance of different constraints, however, can be controlled by degrees of dilation. Important constraints can use dilations corresponding to a small spacing between the dilation contours. A hard constraint might correspond to a physical law that can't be compromised. The corresponding convex sets cannot be dilated. One or more such convex constraints can remain totally undilated while other design constrains can be relaxed through dilation.
	
	Lastly, note that dilated POCS can be interpreted using the idea of  fuzzy convex sets as introduced in Zadeh's seminal paper \cite{Zadeh}. The contours of the fuzzy membership function of a fuzzy convex set are conventional convex sets \cite{marks1995, oh1993} akin to the illustration in Figure~\ref{fig-Dilation}. The dilation search becomes the problem of identifying where the contours of two or more  fuzzy convex sets touch. 
	
    \section*{Acknowledgment}
    This work has been funded by the Army Research Office (Grant No. W911NF-20-2-0270).  The views and opinions expressed do not necessarily represent the opinions of the U.S. Government.

\end{document}